\documentclass[twocolumn,showpacs,preprintnumbers,amsmath,amssymb]{revtex4}
\usepackage{graphicx}

\begin{document}

\title{Enhanced stochastic oscillations in autocatalytic reactions}

\author{Thierry Dauxois}
\affiliation{Laboratoire de Physique, Ecole Normale Sup\'erieure Lyon, 
CNRS, France}
\author{Francesca Di Patti}
\affiliation{CSDC Centro Interdipartimentale per lo Studio di Dinamiche 
Complesse,
University of Florence, Italy and INFN}
\author{Duccio Fanelli}
\affiliation{Dipartimento di Energetica, University of Florence,
Via S. Marta 3, 50139 Florence, Italy and INFN
}
\author{Alan J. McKane}
\affiliation{ 
Theoretical Physics, School of Physics and Astronomy,
University of Manchester, Manchester M13 9PL, United Kingdom
} 

\begin{abstract}
We study a simplified scheme of $k$ coupled autocatalytic reactions, previously
introduced by Togashi and Kaneko. The role of stochastic fluctuations is 
elucidated through the use of the van Kampen system-size expansion and the
results compared with direct stochastic simulations. Regular temporal 
oscillations are predicted to occur for the concentration of the various 
chemical constituents, with an enhanced amplitude resulting from a resonance 
which is induced by the intrinsic graininess of the system. The associated 
power spectra are determined and have a different form depending on the number 
of chemical constituents, $k$. We make detailed comparisons in the two cases 
$k=4$ and $k=8$. Agreement between the theoretical and numerical results for 
the power spectrum are good in both cases. The resulting spectrum is especially
interesting in the $k=8$ system, since it has two peaks, which the system-size 
expansion is still able to reproduce accurately.  
\end{abstract} 


\maketitle
 
\vspace{0.8cm}

\section{Introduction}
\label{intro}
Autocatalytic reactions have long fascinated physicists and chemists because
of their unique features\,\cite{gra85}. A chemical reaction is called 
autocatalytic if one of the reaction products is itself a catalyst for the 
chemical reaction. Part of the reason for the interest in these types of 
reactions stems from the fact that even if only a small amount of the catalyst 
is present, the reaction may start off slowly, but will quickly speed up once 
more catalyst is produced. If the reactant is not replaced, the process will 
again slow down producing the typical sigmoid shape for the concentration of 
the product. All this is for a single chemical reaction, but of greater 
interest is the case of many chemical reactions, where one or more reactions 
produce a catalyst for some of the other reactions. Then the whole collection 
of constituents is called an autocatalytic set\,\cite{jai98}. In addition to 
the interesting properties of autocatalytic sets, there is also an intriguing 
possibility that ``bootstrap'' reactions such as this may have had an 
important role in producing complex or self-replicating molecules required for 
the origin of life on Earth\,\cite{dys85,kau86,sta90,wac90}.

Theoretical studies of the properties of autocatalytic reactions are typically 
of two kinds. In the first, rate equations for the reactions are written down 
and these are either solved numerically or their properties investigated using 
the techniques used in the study of dynamical systems. An alternative is to 
carry out computer simulations of the actual reactions themselves. However 
there is a third possibility: using methods from the theory of stochastic 
processes an analytic approach to the full model (and not just the mean field 
version) is possible. In the last few years this last approach has been used 
for systems which are closely related to autocatalytic reactions, such as 
predator-prey interactions \cite{mck05}, metabolic reactions \cite{mck07}, and 
epidemic models \cite{alo07}. These all show oscillatory behavior in the 
number of individuals or constituents, which arise from feedbacks. These 
oscillations are distinct from the limit cycles found in the rate equations, 
and are purely stochastic in origin. The main tool that is used to analyze 
these systems is the system-size expansion of van Kampen \cite{van07,mck04} 
which gives very good agreement with the simulation results, even for systems 
of a moderate size.

In this paper we apply this technique to the autocatalytic reaction scheme 
studied by Togashi and Kaneko \cite{tog01,tog03}. In most autocatalytic 
reactions there are two types of constituent: the autocatalytic and the
substrate. The number of the latter type are kept constant by continually 
feeding them in, however the former are not injected nor extracted from the 
system. In this sense the system is closed as far as the autocatalytic 
constituents are concerned, but open for the substrate. In the scheme that
Togashi and Kaneko investigate, the reactions are cyclic, with $k$ 
constituents $X_{1},\ldots,X_{k}$ reacting according to 
$X_{i}+X_{i+1} \rightarrow 2 X_{i+1}$ with $X_{k+1} \equiv X_{1}$, 
$i=1,\ldots,k$. The chemicals are assumed to be in a container which is 
well-stirred, but with the possibility of diffusing across the surface of the
container into a particle reservoir.

In their approach Togashi and Kaneko \cite{tog01,tog03} use only computer 
simulation to study this reaction scheme. The analytic techniques we will 
use begin by writing down the master equation for their reaction scheme, and
then studying it through a systematic expansion in $N^{-1/2}$, where $N$ is 
the system size. To leading order one finds the rate equations which appear in
\cite{tog01}, and to next-to-leading order a Langevin equation which describes
the fluctuations about the stable fixed point of the rate equations. From
previous work we expect that (i) this first-order correction will be sufficient
to yield results which are in good agreement with simulation data, (ii) the
large amplitude of the oscillations
can be understood as a resonant effect. One of the strengths of the technique
is that the next-to-leading order corrections give \textit{linear} Langevin
equations which can be analyzed exactly for arbitrary values of $k$.

The outline of the paper is as follows. In Section \ref{govern} we derive the
equations which govern the dynamics of the system, both in the deterministic 
limit and for the fluctuations about this limit. These fluctuations are 
analyzed in Section \ref{analyse} by calculating the power spectra for each 
chemical species $i$. Theoretical predictions are then compared to direct 
simulations for the case $k=4$ and $k=8$. Finally in Section \ref{concl} we 
sum up and discuss possible future work. An Appendix contains the intermediate 
steps required to find the equations given in the main text.  

\section{Governing equations}
\label{govern}

The autocatalytic reaction scheme described in Section \ref{intro} can be 
formulated as
\begin{eqnarray}
X_{i} + X_{i+1} &\stackrel {r_{i+1}}{\rightarrow }& 2X_{i+1}\,, \ 
X_{k+1} \equiv X_{1} \nonumber \\ 
E \stackrel {\alpha_i}{\rightarrow } X_{i} \ 
&;& \  X_{i} \stackrel{\beta_i}{\rightarrow } E\,, \ i=1,\ldots,k\,.
\label{reactions}
\end{eqnarray}
Here $r_{i}, \alpha_{i}$ and $\beta_{i}$ (with $r_{k+1} \equiv r_{1}$), are 
the rates at which the reactions take place and $E$ is the null constituent. 
Such constituents have to be included so that the number of molecules of type 
$X_i$, $n_i$, are all independent. If the size of the system is denoted by $N$,
then $\sum^{k}_{i=1} n_{i} + n_{E} = N$, where $n_E$ is the number of null 
constituents. While $N$ is fixed, $n_E$ may vary as the total number of 
molecules changes with time. In practice, $n_E$ does not explicitly appear 
in the formalism; it is always replaced by $N - \sum^{k}_{i=1} n_{i}$. The
rate constants $\alpha_i$ and $\beta_i$ in Eq.~(\ref{reactions}) represent the 
interactions of the system with the particle reservoir outside the container. 
In effect $\alpha_i$ and $\beta_i$ are the rate at which molecules appear and
disappear from the system, and thus are analogous to birth and death rates.

As an aside, we note that reaction rates which result from a binary encounter
should be scaled by the volume of the system, $V$. That is, 
$r_i \rightarrow r_i/V$. This follows from a straightforward kinetic theory 
argument\,\cite{gil76}. This is an innocent modification as far as this study
is concerned, since it is carried out at constant volume, but it becomes 
crucially important when the volume is allowed to change, as it does in the 
analysis of the phase transition reported in \cite{tog01,tog03}. 

The state of the system is labeled by the set of integers
$\{ n_{1},\ldots,n_{k}\}$ and, under the assumption that the transitions from 
this state to any other only depends on these integers, the system is
Markov and may be described in terms of a master equation. In constructing
the master equation we need to give the transition rates 
$T(\boldsymbol{n}'|\boldsymbol{n})$ from the state $\boldsymbol{n}$ to the  
to the state $\boldsymbol{n}'$, where 
$\boldsymbol{n} \equiv (n_{1},\ldots,n_{k})$. If the system is well-stirred, so
that the probability of a reaction taking place is proportional to its rate and
the number of reactant molecules, then from Eq.~(\ref{reactions}) these 
transition rates are
\begin{eqnarray}
& & T(n_{1},\ldots,n_{i}-1,n_{i+1}+1,\ldots,n_{k}|\boldsymbol{n}) = 
r_{i+1} \frac{n_i}{N} \frac{n_{i+1}}{N}\,, \nonumber \\
& & T(n_{1},\ldots,n_{i}+1,\ldots,n_{k}|\boldsymbol{n}) = 
\alpha_{i} \left( 1 - \frac{\sum^{k}_{j=1} n_j}{N} \right)\,, \nonumber \\
& & T(n_{1},\ldots,n_{i}-1,\ldots,n_{k}|\boldsymbol{n}) = 
\beta_{i} \frac{n_i}{N}\,.
\label{trans_rates}
\end{eqnarray}
The master equation for the probability that the system is in state 
$\boldsymbol{n}$ at time $t$, $P(\boldsymbol{n},t)$, may now be written down:
\begin{eqnarray}
& & \frac{dP(\boldsymbol{n},t)}{dt} = \sum^{k}_{i=1} \left( {\cal E}_{i}
{\cal E}^{-1}_{i+1} - 1 \right) \nonumber \\
& & \times \left[T(n_{1},\ldots,n_{i}-1,n_{i+1}+1,\ldots,n_{k}|\boldsymbol{n})
P(\boldsymbol{n},t) \right] \nonumber \\
& & + \sum^{k}_{i=1} \left( {\cal E}^{-1}_{i} - 1 \right) 
\left[T(n_{1},\ldots,n_{i}+1,\ldots,n_{k}|\boldsymbol{n}) 
P(\boldsymbol{n},t) \right] \nonumber \\
& & + \sum^{k}_{i=1} \left( {\cal E}_{i} - 1 \right) 
\left[T(n_{1},\ldots,n_{i}-1,\ldots,n_{k}|\boldsymbol{n}) 
P(\boldsymbol{n},t) \right] 
\label{master}
\end{eqnarray}
where ${\cal E}^{\pm 1}_{i}$ are the step-operators introduced by van 
Kampen \cite{van07}:
\begin{equation}
{\cal E}^{\pm 1}_{i} f(\boldsymbol{n}) = 
f(n_1,\ldots,n_{i} \pm 1,\ldots,n_k)\,.
\label{step_operators}
\end{equation}

Equations such as (\ref{master}) are difficult to analyze, but if one is 
particularly interested in large or moderately sized values of $N$, then the 
system-size expansion provides an elegant way of encapsulating the essential 
aspects of the model. The key assumption of the method is to write \cite{van07}
\begin{equation}
\frac{n_i}{N} = \phi_{i}(t) + \frac{\xi_{i}(t)}{\sqrt{N}}\,.
\label{key}
\end{equation}
From this relation, $\lim_{N \to \infty} (n_i/N) = \phi_{i}(t)$, the fraction 
of the molecules which are of type $X_i$ at time $t$, within the mean-field
($N \to \infty$) limit. The fluctuations about these are assumed to be 
Gaussian, hence the $1/\sqrt{N}$ in Eq.~(\ref{key}). One of the consequences of
this assumption is that one is looking at a regime sufficiently far from 
boundaries that the probability density functions of the $X_i$ are Gaussian. 
This implies that stochastic extinctions will not be well-described by the 
method, at least to leading order. 

Substituting Eq.~(\ref{key}) into Eq.~(\ref{master}) allows us to expand the
master equation as a power series in $1/\sqrt{N}$. To see this we first note 
that the step operators (\ref{step_operators}) take a particularly simple 
form within the method \cite{van07}
\begin{equation}
{\cal E}^{\pm 1}_{i} = 1 \pm \frac{1}{\sqrt{N}} \frac{\partial }
{\partial \xi_i} + \frac{1}{2N} \frac{\partial^2 }
{\partial \xi_i^2} + \ldots\,.
\label{S_O}
\end{equation}
If we set $P(\boldsymbol{n},t)$ equal to $\Pi(\boldsymbol{\xi},t)$, the
left-hand side of the master equation becomes \cite{van07}
\begin{equation}
\frac{dP(\boldsymbol{n},t)}{dt} = \frac{\partial \Pi(\boldsymbol{\xi},t)}
{\partial t} - \sqrt{N} \sum^{k}_{i=1} 
\frac{\partial \Pi(\boldsymbol{\xi},t)}{\partial  \xi_i} \frac{d\phi_i}{dt}\,.
\label{LHS}
\end{equation}
Substituting Eq.~(\ref{key}) into the right-hand side of the master equation
(\ref{master}), and using the transition rates (\ref{trans_rates}), we may
equate terms of the same order in $1/\sqrt{N}$ on the left- and right-hand 
sides. To leading order this gives (see Appendix \ref{append} for details)
\begin{equation}
\frac{d\phi_i}{d\tau} = \left( r_{i} \phi_{i-1} - r_{i+1} \phi_{i+1} \right) 
\phi_{i} + \alpha_{i} \left( 1 - \sum^{k}_{j=1} \phi_{j} \right) 
- \beta_{i} \phi_{i}\,,
\label{deterministic}
\end{equation}
where $\tau$ is a rescaled time: $\tau=t/N$. At next order one finds a 
Langevin equation:
\begin{equation}
\frac{d\xi_i}{d\tau} = \sum^{k}_{j=1} M_{ij} \xi_{j}(\tau) + \eta_{i}(\tau)\,,
\label{Langevin}
\end{equation}
where $M$ is a $k \times k$ matrix which can be found from Eqs.~(\ref{A_i}) and
(\ref{AandM}), and $\eta_{i}$ is a Gaussian white noise with zero mean and 
correlator
\begin{equation}
\langle \eta_{i}(\tau) \eta_{j}(\tau') \rangle = B_{ij} 
\delta \left( \tau - \tau' \right)\,,
\label{correlator}
\end{equation}
and $B_{ij}$ is another $k \times k$ matrix given by Eq.~(\ref{B_ij}).
 
The first equation, Eq.~(\ref{deterministic}), is a deterministic equation 
for the fraction of molecules which are of type $i$. It agrees with that of 
Togashi and Kaneko \cite{tog01}, if one takes into account that their equations
are for concentrations and so contain the (constant) concentrations of the
species in the reservoir. There is also an additional term $\sum_{j} \phi_j$ in
Eq.~(\ref{deterministic}), which is typically present when mean-field equations
are derived in systems with a fixed size, but not in the phenomenologically 
postulated form. For small concentrations it will not be important, but 
clearly it will have an effect as the ceiling on particle numbers is felt, 
reducing the number of molecules entering the container from the reservoir, as
it should. The second equation, Eq.~(\ref{Langevin}), is a stochastic 
differential equation for the deviation from these mean-field values. It is 
the analysis of these two equations that allow us to describe the stochastic 
aspects of the autocatalytic reactions in a quantitative way. 

\section{Analysis of the fluctuations}
\label{analyse}

In their numerical studies, Togashi and Kaneko \cite{tog01,tog03} looked at the
simplest case of the model where the rates $r_i, \alpha_i$ and $\beta_i$ were
the same for all chemical species. To illustrate our method we will do the 
same, and so from now on we will drop the subscript $i$ on these constants, but
it should be clear that our analysis also applies to the general situation 
where they are different for each species. With this choice, the deterministic 
equations (\ref{deterministic}) have a single fixed point:
\begin{equation}
\phi^{*} = \frac{\alpha}{\beta + k\alpha}\,,
\label{FP}
\end{equation}
where the asterisk denotes the fixed point value. 

If $N$ is so large that the fluctuations are completely negligible, then the 
system tends towards a state where the fractions of the chemical species in
the system are equal, and given by Eq.~(\ref{FP}), and stays there. Of course,
if $N$ is finite this is no longer the case and there are fluctuations about 
this stationary state --- and as we will see these can be significant even if 
$N$ is quite large. Since these fluctuations are expected to be oscillatory, 
we begin their analysis by taking the Fourier transform of Eq.~(\ref{Langevin})
to find
\begin{equation}
\sum^{k}_{j=1} \left( -i\omega \delta_{ij} - M_{ij} \right) \tilde{\xi}_{j} 
(\omega) = \tilde{\eta}_{i}(\omega)\,,
\label{FT_Lang}
\end{equation}
where the $\tilde{f}$ denotes the Fourier transform of the function $f$.
Defining the matrix $-i\omega \delta_{ij} - M_{ij}$ to be $\Phi_{ij}(\omega)$,
the solution to Eq.~(\ref{FT_Lang}) is
\begin{equation}
\tilde{\xi}_{i}(\omega) = \sum^{k}_{j=1} \Phi^{-1}_{ij}(\omega) 
\tilde{\eta}_{j}(\omega)\,.
\label{solution}
\end{equation}
  
To identify the dominant frequency of the oscillatory behavior, we compute the
power spectrum for the $i$th species, $P_{i}(\omega)$, from 
Eq.~(\ref{solution}):
\begin{equation}
P_{i}(\omega) \equiv \left\langle | \tilde{\xi}(\omega) |^{2} \right\rangle
= \sum^{k}_{j=1} \sum^{k}_{l=1} \Phi^{-1}_{ij}(\omega)B_{jl}
\left( \Phi^{\dag} \right)^{-1}_{li}(\omega)\,,
\label{PS_defn}
\end{equation}
Since $\Phi = -i\omega I - M$, where $I$ is the $k \times k$ unit matrix, and
since $M$ and $B$ are independent of $\omega$, the structure of $P_{i}(\omega)$
is that of a polynomial of order $2k$ divided by another polynomial of degree 
$2k$. The explicit form of the denominator is $|\det \Phi (\omega) |^{2}$.

From previous investigations of fluctuations of a similar 
kind \cite{mck05,alo07,mck07}, we expect that the fluctuations about the 
stationary state (\ref{FP}) will be enhanced by a resonant effect: for values
of $\omega$ for which $|\det \Phi (\omega) |$ is a minimum, the power spectra
will show peaks which correspond to larger than expected fluctuations at that 
frequency. This effect was first conjectured by Bartlett \cite{bar57} in the
context of the modeling of measles epidemics, and later elaborated upon by 
Nisbet and Gurney \cite{nis82}, who called these stochastically induced cycles,
quasi-cycles. However it is only in the last few years that explicit 
calculations within the system-size expansion have been carried out and a 
quantitative understanding of the phenomenon has emerged \cite{mck05}.

To understand the analytic structure of the power spectra, we begin by 
supposing that we can neglect the effects of the numerator on the right-hand 
side of Eq.~(\ref{PS_defn}), and simply determine the dominant frequency 
by looking for the value which minimizes $|\det \Phi (\omega) |$. The effect
of the numerator will be to shift this frequency; we are assuming as a first 
approximation that this shift will be small, as indeed it has been found to 
be in some cases \cite{mck05}. If $\lambda_{j}$ are the eigenvalues of $M$,
then the denominator of the expression for the power spectra may be written as
\begin{equation}
|\det \Phi (\omega) |^{2} = \prod^{k}_{j=1} 
\left( -i\omega - \lambda_{j} \right) \left( i\omega 
- \lambda^{*}_{j} \right)\,.
\label{det_Phi}
\end{equation}
Since $M$ is real, the $\lambda_{j}$ will be real or come in complex conjugate 
pairs, so that the products in Eq.~(\ref{det_Phi}) has one of two forms:
\begin{itemize}
\item[(i)]If $\lambda$ is real, the two factors involving this eigenvalue give 
$(\omega^{2}+\lambda^{2})$.
\item[(ii)]If $\lambda$ is complex: $\lambda=\lambda_{R}+i\lambda_{I}$, the 
four terms involving $\lambda$ and $\lambda^{*}$ give
\begin{equation}
\left| \omega^{2} + \left( \lambda^{2}_{R} - \lambda^{2}_{I} \right) 
+ 2i \lambda_{R}\lambda_{I} \right|^{2}\,.
\label{omega_sq_term}
\end{equation}
\end{itemize}
The resonant effect has its origin in the structure of the factor coming from 
the complex eigenvalues shown in the expression (\ref{omega_sq_term}). It is
smallest, and so gives the largest contribution when it is in the denominator, 
for frequencies which satisfy
\begin{equation}
\omega^{2}_{c}=\lambda^{2}_{I}-\lambda^{2}_{R}\,.
\label{peak_cond}
\end{equation}
If there are several pairs of complex eigenvalues and their conjugates, the 
largest contribution should come from the pair for which 
$\lambda_{R}\lambda_{I}$ is smallest. Clearly this will only be approximately 
true since, not only are we neglecting the numerator, but also the factors 
$(\omega^{2}+\lambda^{2})$ coming from real eigenvalues, as well as those 
coming from other complex conjugate pairs. However, as we will now see by
looking at two specific cases, $k=4$ and $k=8$, these approximations appear to
be remarkably good.

We study the cases $k=4$ and $k=8$ because they are the smallest even values
of $k$ for which one complex conjugate pair and two distinct complex conjugate
pairs, respectively, exist (there are two complex conjugate pairs for $k=6$,
but they are equal, and three for $k=8$, but two of these are equal). We
therefore expect to see one peak in the power spectra when $k=4$ and two 
when $k=8$. Our analysis, and the accuracy of our approximations, can be
directly checked by numerical simulation of the chemical reaction system
(\ref{reactions}) by use of the Gillespie algorithm \cite{gil76,gil77}. This
produces realizations of the stochastic dynamics which are equivalent to 
those found from the master equation (\ref{master}). Averaging over many of 
these realizations gives us power spectra after Fourier transformation, which 
are exact to a given numerical accuracy. We now investigate the two cases $k=4$
and $k=8$ in more detail.

\subsection{Power spectra when $k=4$}
\label{k_eq_4}

\begin{figure}[b]
\vspace{2truecm}
\includegraphics[width=8cm]{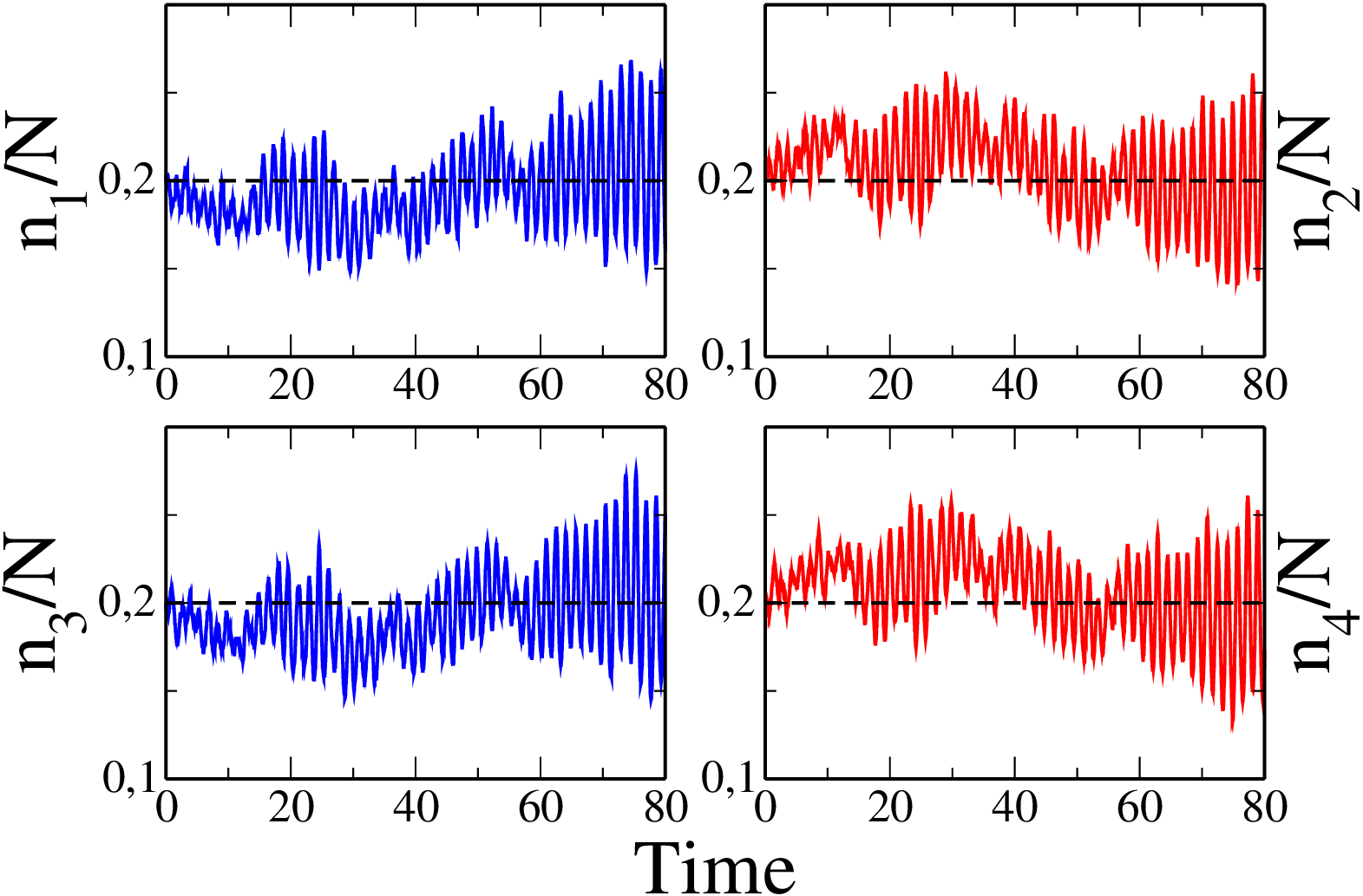}
\caption{(Color online) Time evolution of selected species, $i=1,2,3,4$ in
clockwise order for the case $k=4$. Here $r=10$, $\alpha=\beta=1/64$, $N=8192$.
The dashed line indicates the mean field solution. The species display a clear 
oscillatory trend about their mean field values. A paired synchronization, 
$(1,3)$ vs.\,$(2,4)$ rich states, is also visible, as already observed in 
\cite{tog01,tog03}.}
\label{figure1}
\end{figure}

The time evolution of the species is depicted Figure~\ref{figure1}. This 
clearly displays large oscillations which we aim to investigate analytically. 
Before beginning this analysis, we observe that species $1$, $3$ (odd) and  
$2$, $4$ (even) are paired together and move up and down from the reference 
mean-field level in a synchronized fashion. This fact was already recognized 
in \cite{tog01,tog03} and shown to drive successive switches between the 1-3 
or 2-4 rich states, close to the absorbing boundary, i.e. when a small number 
of molecules is simulated. The rate at which the changes occur is controlled 
by the diffusion parameter. However, the details of the transitions stem from  
a purely dynamical effect which cannot be captured within the perturbative 
analysis developed here. 

\begin{figure}[t]
\vspace{2truecm}
\includegraphics[width=7cm]{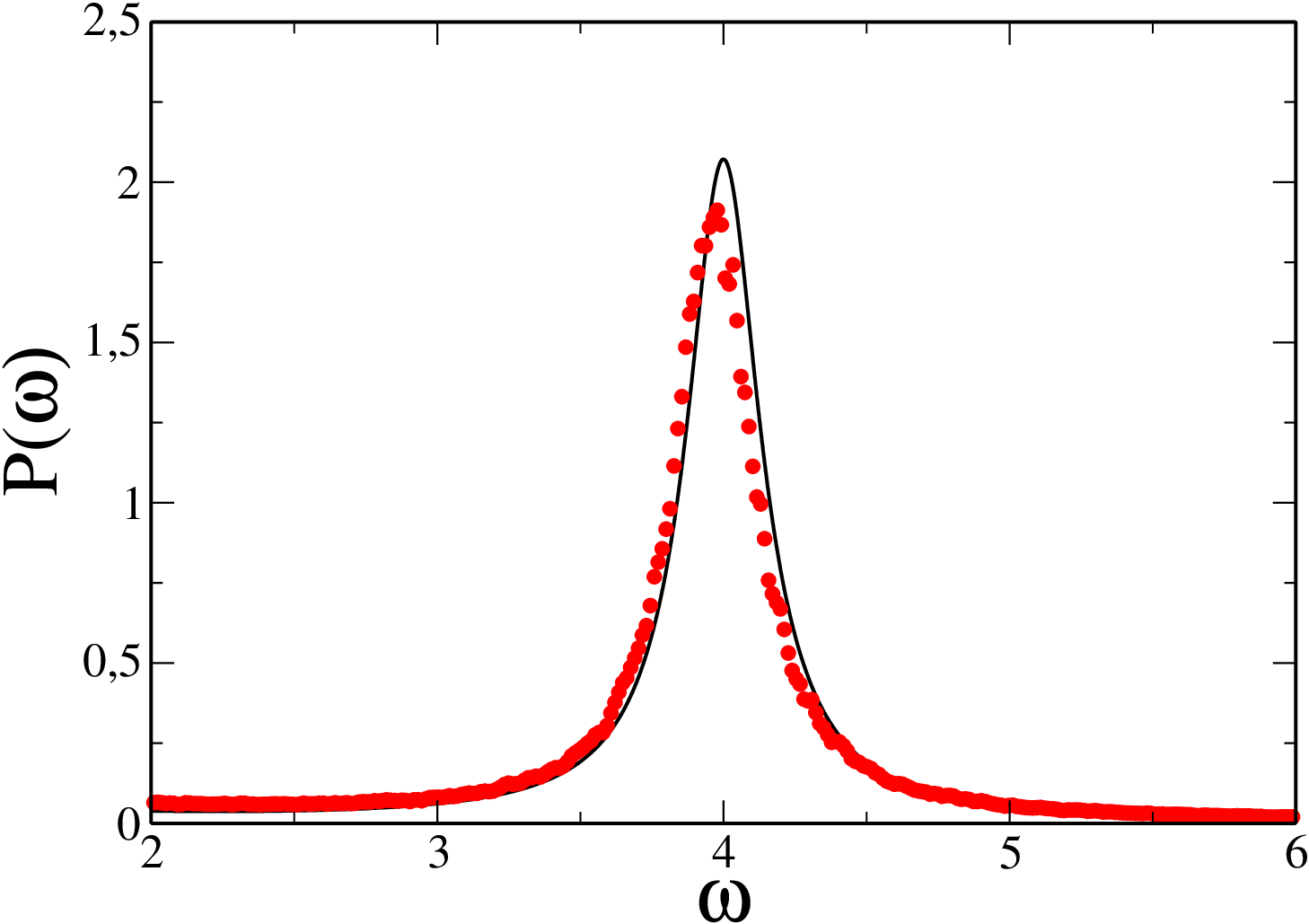}
\caption{(Color online) Power spectrum of species $i=2$ when $k=4$. The 
analytical curve is shown as a solid line and the simulation (average over 
500 independent realizations) as symbols. Here 
$r=10$, $\alpha= \beta=10/64$, $N=5000$.}
\label{figure2}
\end{figure}

Let us now turn to analytically characterizing the aforementioned oscillatory 
regime. To this end we begin by determining the eigenvalues of the $M$ matrix. 
From Eq.~(\ref{eigen2}), these are
\begin{eqnarray}
\lambda_{0} &=& \beta + 4\alpha\,, \ \ \lambda_{2} = \beta\,, \nonumber \\
\lambda_{1} &=& \beta + 2ir\phi^{*}\,, \ \ \lambda_{3} = \lambda_{1}^{*}\,.
\label{k_4_eigen}
\end{eqnarray}
Within the approximations we have discussed, we would expect that there should
be a single peak in the power spectrum for any one of the chemical species
at a frequency given by (see Eq.~(\ref{peak_cond}))
\begin{equation}
\omega^{2}_{c} = 4r^{2}\left( \phi^{*} \right)^{2} - \beta^2 
= \frac{4r^{2}\alpha^2}{\left( \beta + 4\alpha \right)^2} - \beta^{2}\,.
\label{peak_cond_4}
\end{equation}
In Fig.~\ref{figure2} we show the power spectrum (for the chemical species 
$i=2$) found by averaging over 500 realizations from the Gillespie algorithm,
together with that found from Eq.~(\ref{PS_defn}) using the matrices $B$ and
$M$ given in the Appendix. The good agreement between the simulation results
and those found from applying the system-size expansion, shows that the 
method works well for $N=5000$. The parameters used in this case were $r=10$ 
and $\alpha=\beta=1/64$, which gives a value of $\omega_{c} \approx 4$ from
Eq.~(\ref{peak_cond_4}). From Fig.~\ref{figure2} we see this is a surprising 
good estimate for the position of the peak, given the significant frequency 
dependence which we have neglected to obtain the estimate (\ref{peak_cond}).

\begin{figure}[t]
\vspace{2truecm}
\includegraphics[width=7cm]{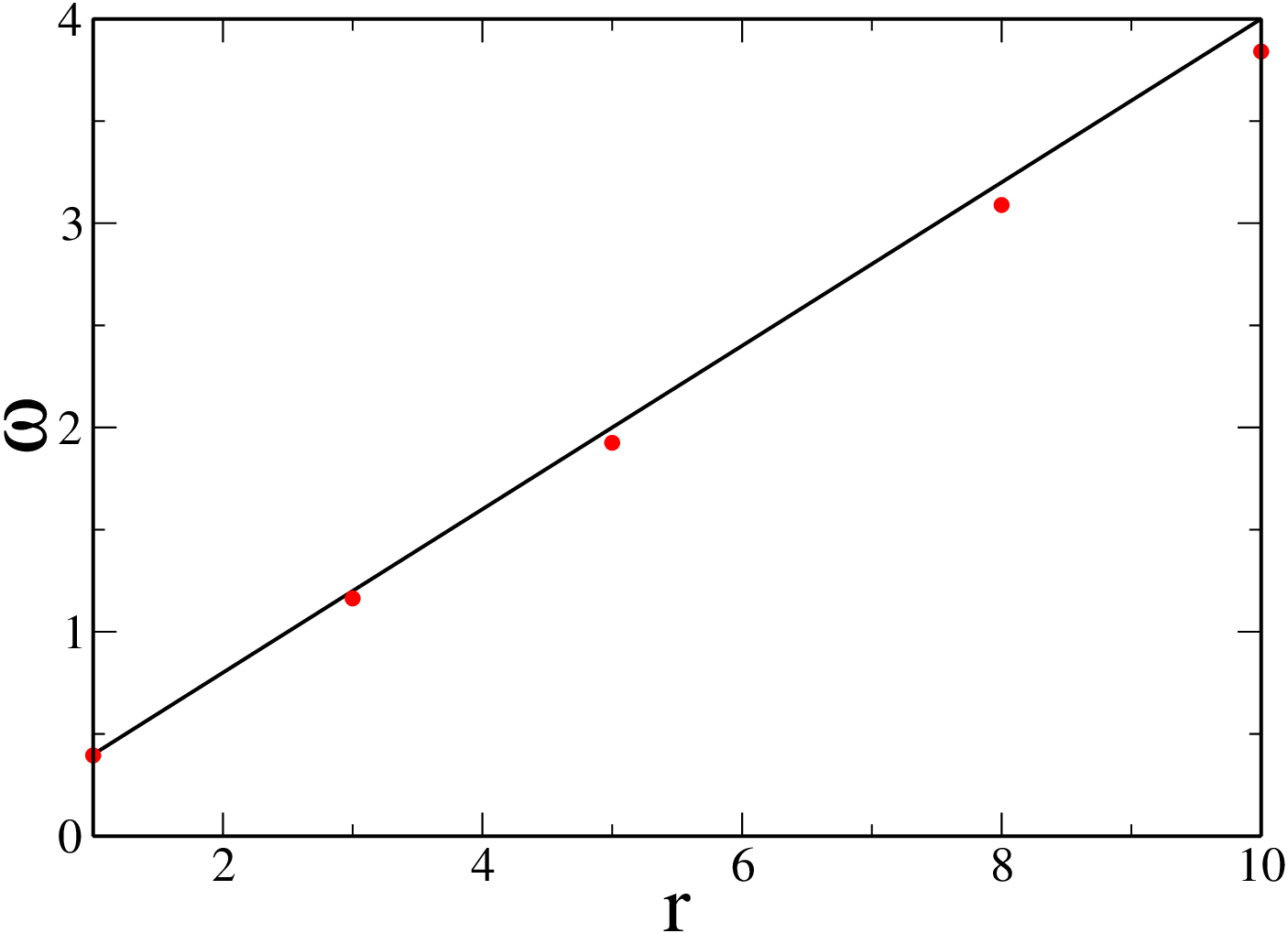}
\caption{(Color online) The position of the peak in the power spectrum (for
species $i=2$ when $k=4$) plotted as function of the rate constant $r$. 
Symbols refers to the stochastic simulations, while the solid line shows the 
analytical prediction. Here $\alpha=\beta=1/64$, $N=5000$}
\label{figure3}
\end{figure}

Another check of the accuracy of these approximations, and so of 
Eq.~(\ref{peak_cond}), is to imagine increasing the parameter $\beta$ at fixed
$r$ and $\alpha$, and asking when $\omega^{2}_{c}$ will become zero, and so 
at what frequency will the peak in the power spectra disappear. From 
Eq.~(\ref{peak_cond_4}) we estimate this to be
\begin{equation}
\beta \sim \frac{2r\alpha}{\beta} \ \ {\rm or} \ \ 
\beta \sim \sqrt{2r\alpha}\,,
\label{estimate_k4}
\end{equation}
which equals $0.56$ for the values of $r$ and $\alpha$ used in 
Fig.~\ref{figure2}. Once again this agrees well with the full spectrum which 
predicts the peak to disappear at about the same value. As a final check, we 
measure the position of the peak from a set of simulations run at different 
values of $r$. Direct measurements (symbols) are compared to the theory (solid 
line) in Figure \ref{figure3} and are in good quantitative agreement. Again, 
we recall that adjusting the rate $r$ can be equivalently seen as modifying 
the volume of the system, which is the setting investigated 
in \cite{tog01,tog03}.  

%%%%%%%%%%%%%%%%%%%%%%%%%%%%%%%%%%%%%%%%%%%%%%%%%%%%%%%%%%%%%%%%%%%%%%%%%%%
%%%%%%%%%%%%%%%%%%%%%%%%%%%%%%%%%%%%%%%%%%%%%%%%%%%%%%%%%%%%%%%%%%%%%%%%%%%

%%%%%%%%%%%%%%%%%%%%%%%%%%%%%%%%%%%%%%%%%%%%%%%%%%%%%%%%%%%%%%%%%%%%%%%%%%%
%%%%%%%%%%%%%%%%%%%%%%%%%%%%%%%%%%%%%%%%%%%%%%%%%%%%%%%%%%%%%%%%%%%%%%%%%%%

%%%%%%%%%%%%%%%%%%%%%%%%%%%%%%%%%%%%%%%%%%%%%%%%%%%%%%%%%%%%%%%%%%%%%%%%
%%%%%%%%%%%%%%%%%%%%%%%%%%%%%%%%%%%%%%%%%%%%%%%%%%%%%%%%%%%%%%%%%%%%%%%%

%%%%%%%%%%%%%%%%%%%%%%%%%%%%%%%%%%%%%%%%%%%%%%%%%%%%%%%%%%%%%%%%%%%%%%%%%%
%%%%%%%%%%%%%%%%%%%%%%%%%%%%%%%%%%%%%%%%%%%%%%%%%%%%%%%%%%%%%%%%%%%%%%%%%%

%%%%%%%%%%%%%%%%%%%%%%%%%%%%%%%%%%%%%%%%%%%%%%%%%%%%%%%%%%%%%%%%%%%%%%%%%%
%%%%%%%%%%%%%%%%%%%%%%%%%%%%%%%%%%%%%%%%%%%%%%%%%%%%%%%%%%%%%%%%%%%%%%%%%%

%%%%%%%%%%%%%%%%%%%%%%%%%%%%%%%%%%%%%%%%%%%%%%%%%%%%%%%%%%%%%%%%%%%%%%%%%%%%% 
%%%%%%%%%%%%%%%%%%%%%%%%%%%%%%%%%%%%%%%%%%%%%%%%%%%%%%%%%%%%%%%%%%%%%%%%%%%%%

\subsection{Power spectra when $k=8$}
\label{k_eq_8}

\begin{figure}[t]
\includegraphics[width=7cm]{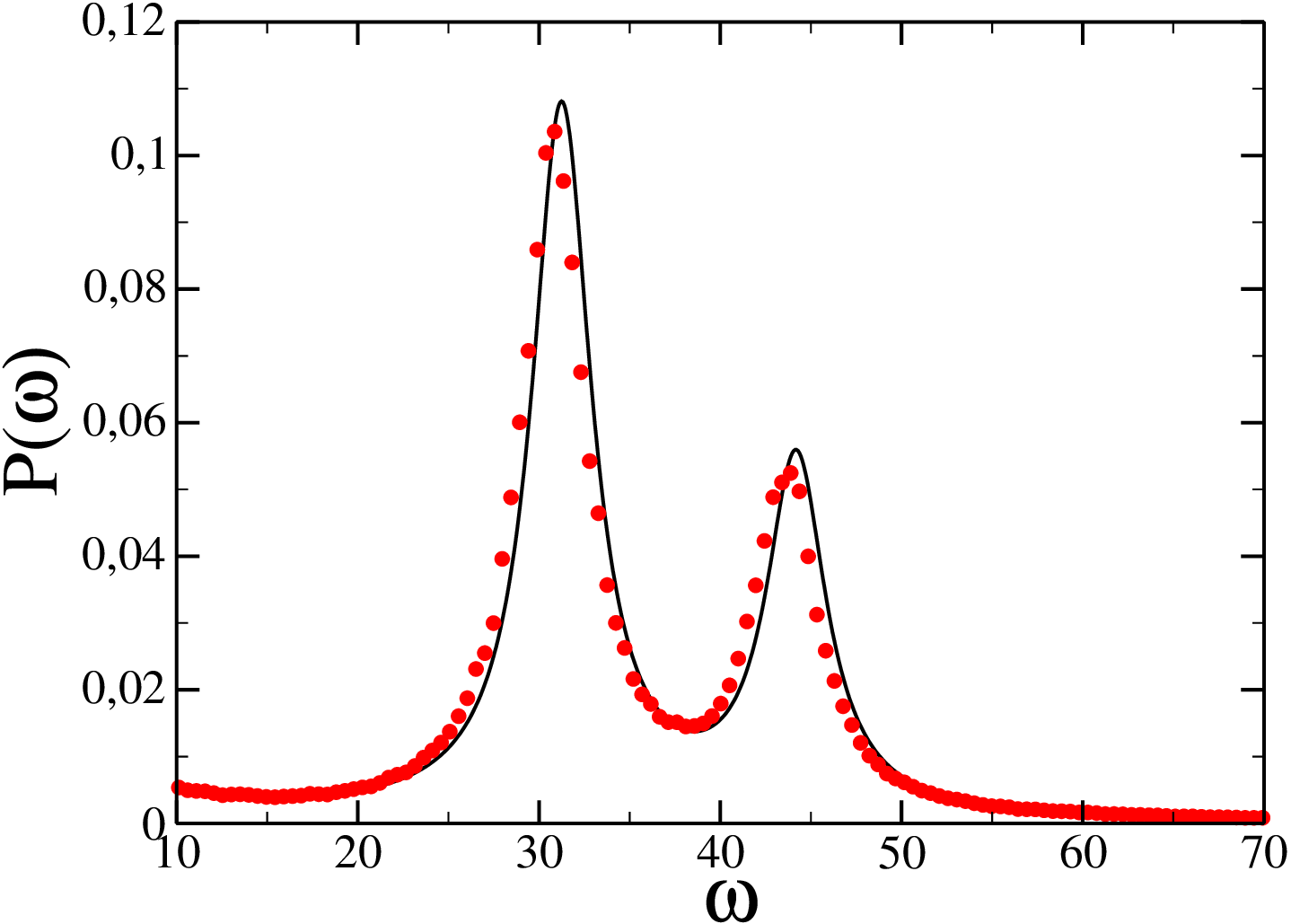}
\caption{(Color online) Power spectrum of the time series for species $i=2$ 
when $k=8$. The analytical result (solid line) is superimposed onto the 
simulations (symbols), averaged over 500 independent realizations.  Here 
$r=200$, $\alpha=1.9$, $\beta=2$, $N=7000$.}
\label{figure4}
\end{figure}

From Eq.~(\ref{eigen2}), the eigenvalues of the $M$ matrix are
\begin{eqnarray}
\lambda_{0} &=& \beta + 8\alpha\,, \ \ \lambda_{4} = \beta\,, \nonumber \\
\lambda_{1} &=& \beta + \sqrt{2}ir\phi^{*}\,, \ \ 
\lambda_{7} = \lambda_{1}^{*}\,, \nonumber \\
\lambda_{2} &=& \beta + 2ir\phi^{*}\,, \ \ \lambda_{6} = \lambda_{2}^{*}\,,
\nonumber \\
\lambda_{3} &=& \beta + \sqrt{2}ir\phi^{*}\,, \ \ \lambda_{5} = 
\lambda_{3}^{*}\,.
\label{k_8_eigen}
\end{eqnarray}
Since there are two distinct complex conjugate pairs we would expect to find 
two peaks in the power spectra, one at $\omega^2_c = 2r^2(\phi^*)^2 - \beta^2$ 
and the other at $\omega^2_c = 4r^2(\phi^*)^2 - \beta^2$. For small $\beta$,
one peak will be at a frequency $\sqrt{2}$ times the other. We would
also expect that the peak at lower frequency would be larger than the one
at higher frequency, since $\lambda_{R}\lambda_{I}$ is smaller for the former. 
That is, the pole in the power spectra in the complex frequency squared
plane is nearer to the real axis for the peak at lower frequency, and so should
have a bigger effect. So, in summary, our approximations indicate that the
peaks in the power spectra should be given by
\begin{equation}
\omega^{2}_{c1} = \frac{2r^{2}\alpha^2}{\left( \beta + 8\alpha \right)^2} 
- \beta^{2}\,, \ \ \omega^{2}_{c2} = \frac{4r^{2}\alpha^2}
{\left( \beta + 8\alpha \right)^2} - \beta^{2}\,, 
\label{peak_cond_8}
\end{equation}
with the peak at $\omega=\omega_{c1}$ larger than the one at 
$\omega=\omega_{c2}$. The results of plotting the full spectrum found from
Eq.~(\ref{PS_defn}) and simulation results are shown in Fig.~\ref{figure4}
for $r=200, \alpha=1.9$ and $\beta=2$. This corresponds to peaks at
$\omega=31.2$ and $\omega=44.14$, according to Eqs.~(\ref{peak_cond_8}), which
once again agrees very well the the results displayed in the figure, as does 
the prediction that the peak nearest the origin should be the largest. 

%%%%%%%%%%%%%%%%%%%%%%%%%%%%%%%%%%%%%%%%%%%%%%%%%%%%%%%%%%%%%%%%%%%%%%%
%%%%%%%%%%%%%%%%%%%%%%%%%%%%%%%%%%%%%%%%%%%%%%%%%%%%%%%%%%%%%%%%%%%%%%%

%%%%%%%%%%%%%%%%%%%%%%%%%%%%%%%%%%%%%%%%%%%%%%%%%%%%%%%%%%%%%%%%%%%%%%%%
%%%%%%%%%%%%%%%%%%%%%%%%%%%%%%%%%%%%%%%%%%%%%%%%%%%%%%%%%%%%%%%%%%%%%%%%

\section{Conclusion}
\label{concl}

Auto-catalytic networks are central in many different contexts and play an 
important role in intracellular biochemical reaction schemes. In this latter 
scenario, species are confined in a closed volume, delimited by the cellular 
membrane. Low concentration can occasionally develop resulting from the 
complex mutual interaction between microscopic actors. Under such conditions, 
fluctuations matter and the effects of the intrinsic discreteness need to be 
properly accounted for. In other words, continuous kinetic equations prove 
inadequate, finite size corrections becoming significant. These aspects were 
numerically substantiated by Togashi and Kaneko\,\cite{tog01,tog03} within the 
framework of a simplified system of $k$ coupled autocatalytic reactions. 

In this paper we have taken this forward by studying analytically the 
associated master equation via a systematic expansion in power of $N^{-1/2}$, 
where $N$ is the system size. To leading order, the mean-field rate equations 
are recovered, while higher order corrections enable us to explain the large 
amplitude of the oscillations as detected in direct simulations. Importantly, 
the calculation applies to arbitrary values of $k$. For $k=4$ a peak in the 
power spectrum is found, while for $k=8$ two peaks develop. To the best of our 
knowledge, this is the first time that a double-peaked power spectrum has been
predicted to emerge as a resonant effect, within a van Kampen type of analysis.
In both cases, theory and simulations agree well thus confirming the 
importance of finite $N$ contributions. Possible extensions of the present 
work include taking spatial variations into account. This could yield spatial 
oscillations in the species concentration, which would again be driven by the
discreteness of the system components.

\begin{acknowledgments}
We wish to thank K. Kaneko for useful correspondence.
AJM wishes the thank the EPSRC (UK) for financial support under grant 
GR/T11784/01.

\end{acknowledgments}

\appendix

\section{The finite $N$ expansion}
\label{append}
In this appendix we will give the intermediate steps of the calculation using 
the system size expansion, starting with the master equation (\ref{master}) and
ending with the results (\ref{deterministic})-(\ref{correlator}). We will
also give the explicit expressions for the matrices $M$ and $B$.

Applying the ansatz (\ref{key}) to the right-hand side of Eq.~(\ref{master}),
the step-operators (\ref{step_operators}) take the form (\ref{S_O}), the
$n_i$ in the transition rates (\ref{trans_rates}) are replaced by $\phi_i$ and
$\xi_i$ using Eq.~(\ref{key}) and $P(\boldsymbol{n},t)$ becomes 
$\Pi(\boldsymbol{\xi},t)$. This yields the following terms:
\begin{itemize}
\item[(a)] Terms of order $N^{-1/2}$:
\begin{eqnarray}
& & \sum^{k}_{i=1} \left\{ r_{i+1}\phi_{i}\phi_{i+1} 
\left[ \frac{\partial }{\partial \xi_i} 
- \frac{\partial }{\partial \xi_{i+1}} \right] \right. \nonumber \\
& & \left. - \alpha_{i} \left( 1 - \sum^{k}_{j=1} \phi_{j} \right)
\frac{\partial }{\partial \xi_i} 
+ \beta_{i}\phi_{i}\frac{\partial }{\partial \xi_i} \right\} 
\Pi(\boldsymbol{\xi},t)\,.
\label{explicit1}
\end{eqnarray}

\item[(b)] Terms of order $N^{-1}$ and involving first order derivatives:
\begin{eqnarray}
& & \sum^{k}_{i=1} \left\{ r_{i+1}\phi_{i} \frac{\partial }{\partial \xi_i}
\xi_{i+1} + r_{i+1}\phi_{i+1} \frac{\partial }{\partial \xi_i}\xi_{i} 
- r_{i+1}\phi_{i} \frac{\partial }{\partial \xi_{i+1}}\xi_{i+1} \right.
\nonumber \\
& & \left. - r_{i+1}\phi_{i+1} \frac{\partial }{\partial \xi_{i+1}}\xi_{i} 
+ \alpha_{i} \sum^{k}_{j=1} \frac{\partial }{\partial \xi_i} \xi_{j} 
+ \beta_{i} \frac{\partial }{\partial \xi_i} \xi_{i} \right\} 
\Pi(\boldsymbol{\xi},t)\,. \nonumber \\
\label{explicit2}
\end{eqnarray}

\item[(c)] Terms of order $N^{-1}$ and involving second order derivatives:
\begin{eqnarray}
& & \frac{1}{2} \sum^{k}_{i=1} \left\{ r_{i+1}\phi_{i}\phi_{i+1} \left[ 
\frac{\partial^2 }{\partial \xi^2_i} + \frac{\partial^2 }{\partial \xi^2_{i+1}}
- 2 \frac{\partial^2 }{\partial \xi_i \partial \xi_{i+1}} \right] \right.
\nonumber \\
& & \left. + \alpha_{i} \left( 1 - \sum^{k}_{j=1} \phi_{j} \right) 
\frac{\partial^2 }{\partial \xi^2_i} + \beta_{i}\phi_{i} 
\frac{\partial^2 }{\partial \xi^2_i} \right\} \Pi(\boldsymbol{\xi},t)\,.
\label{explicit3}
\end{eqnarray}
\end{itemize}
 Introducing $\tau=t/N$, the terms of order $N^{-1/2}$ in Eq.~(\ref{explicit1})
may be identified with the second term on the right-hand side of 
Eq.~(\ref{LHS}). This gives the $N$ deterministic 
equations (\ref{deterministic}). The terms of order $N^{-1}$ in 
Eqs.~(\ref{explicit2}) and (\ref{explicit3}), are now identified with the 
remaining term on the right-hand side of Eq.~(\ref{LHS}). This resulting 
equation is a Fokker-Planck equation:
\begin{equation}
\frac{\partial \Pi}{\partial \tau} = - \sum_{i}\,\frac{\partial}
{\partial \xi_i} \left[ A_{i} (\boldsymbol{\xi})\,\Pi \right] + 
\frac{1}{2} \sum_{i,j}\,B_{ij} \frac{\partial^{2} \Pi}{\partial \xi_{i} 
\partial \xi_{j}}\,.
\label{FP_eqn}
\end{equation}
From Eq.~(\ref{explicit2}) we see that the $A_{i}(\boldsymbol{\xi})$ are 
linear functions of the $\xi_j$ and from Eq.~(\ref{explicit3}) that the 
$B_{ij}$ are independent of them. Explicitly:
\begin{eqnarray}
A_{i}(\boldsymbol{\xi}) &=& \left( r_{i}\phi_{i-1} - 
r_{i+1}\phi_{i+1} \right)\xi_{i} + r_{i}\phi_{i}\xi_{i-1} \nonumber \\
&-& r_{i+1}\phi_{i}\xi_{i+1} - \alpha_{i}\sum^{k}_{j=1} \xi_{j} - 
\beta_{i}\xi_{i}\,,
\label{A_i}
\end{eqnarray}
and
\begin{equation}
B_{ij} = \left\{ \begin{array}{ll} 
-r_{i} \phi_{i-1}\phi_{i}, & \mbox{\ if $j=i-1$} \\ 
r_{i+1} \phi_{i}\phi_{i+1} + r_{i}\phi_{i}\phi_{i-1} \\
+ \alpha_{i} \left( 1 - \sum^{k}_{j=1} \phi_{j} \right) + \beta_{i}\phi_{i}, 
& \mbox{\ if $j=i$} \\ 
-r_{i+1} \phi_{i}\phi_{i+1}. & \mbox{\ if $j=i+1$}
\end{array} \right.
\label{B_ij}
\end{equation}
In Eqs.~(\ref{A_i}) and (\ref{B_ij}), $\phi_{k+1} \equiv \phi_{1}$ and 
$\xi_{k+1} \equiv \xi_{1}$, which follows from the cyclic nature of the model. 

Since the $A_{i}(\boldsymbol{\xi})$ are linear functions of the $\xi_j$ we 
may write them as
\begin{equation}
A_{i} (\boldsymbol{\xi}) = \sum^{k}_{j=1} M_{ij} \xi_{j}\,.
\label{AandM}
\end{equation}
This means that the probability distribution at next-to-leading order,
$\Pi(\boldsymbol{\xi},\tau)$, is completely determined by the two $k \times k$ 
matrices $M$ and $B$, whose elements are independent of the $\xi_j$, and 
only functions of the $\phi_j$. For our purposes, where we need to Fourier 
analyze the fluctuations, it is more convenient not to use the formulation 
in which the fluctuations are described by a Fokker-Planck equation, but 
rather in terms of Langevin equations. The Fokker-Planck 
equation (\ref{FP_eqn}) is completely equivalent to the Langevin 
equation (\ref{Langevin}) with the 
correlator (\ref{correlator}) \cite{gar04,ris89}, and it is this formalism 
that we will use. 

In principle the matrices $M$ and $B$ are time dependent, since $\phi_j$ is. 
However, in practice we are interested in fluctuations about the stationary 
state, and so we are only interested in the values that the $\phi_j$ take on 
at late times. Furthermore, in Section \ref{analyse} we studied the simple case
$r_{i}=r, \alpha_{i}=\alpha$ and $\beta_{i}=\beta$, for which the relevant 
value of the $\phi_j$ is given by Eq.~(\ref{FP}). With these assumptions $M$
and $B$ are given by
\begin{equation}
M =
\begin{bmatrix}
m_{0} & m_{1} & m_{2} & m_{2} \dots & m_{2} & m_{3} \\
m_{3} & m_{0} & m_{1} & m_{2} \dots & m_{2} & m_{2} \\
m_{2} & m_{3} & m_{0} & m_{1} \dots & m_{2} & m_{2} \\
m_{2} & m_{2} & m_{3} & m_{0} \dots & m_{2} & m_{2} \\
\hdotsfor[2.0]{4}\\
m_{2} & m_{2} & m_{2} & m_{2} \dots & m_{0} & m_{1} \\
m_{1} & m_{2} & m_{2} & m_{2} \dots & m_{3} & m_{0} \\
\end{bmatrix}\,,
\end{equation}
where 
\begin{equation}
m_{0}= - \alpha - \beta\,, m_{1}=  -\alpha - r\phi^{*}\,, m_{2}=-\alpha\,,
m_{3}=-\alpha + r\phi^{*}\,,
\label{ms}
\end{equation}
and
\begin{equation}
B =
\begin{bmatrix}
b_{0} & b_{1} & 0 \dots & 0 & b_{1} \\
b_{1} & b_{0} & b_{1} \dots & 0 & 0 \\
0 & b_{1} & b_{0} \dots & 0 & 0 \\
\hdotsfor[2.0]{4}\\
0 & 0 & 0 \dots & b_{0} & b_{1} \\
b_{1} & 0 & 0 \dots & b_{1} & b_{0} \\
\end{bmatrix}\,,
\end{equation}
where
\begin{equation}
b_{0}=2 r(\phi^{*})^2+\beta \phi^{*}+\alpha (1-k \phi^{*}) \,, 
b_{1}=-r(\phi^{*})^{2}\,.
\label{bs}
\end{equation}

We note that $M$ is a circulant matrix \cite{bel70}, and therefore its 
eigenvalues are given by
\begin{equation}
\lambda_{\ell} = \sum^{k}_{j=1} m_{1j}\,e^{(2\pi i (j-1)\ell)/k}\,, \ \
\ell = 0,1,\ldots,k-1\,,
\label{eigenvalues}
\end{equation}   
where $m_{1j}$ is the element of $M$ in the first row and $j$th column. In
fact, $M$ is not the most general form of circulant matrix; $(k-3)$ entries
in each row are equal (to $m_2$). This leads to a simplified form for the 
eigenvalues:
\begin{eqnarray}
\lambda_{\ell} &=& m_{0} + m_{1}\,e^{2\pi i \ell/k} + 
m_{3}\,e^{-2\pi i \ell/k} + m_{2}\,\sum^{k-2}_{j=2} e^{2\pi i j\ell/k} 
\nonumber \\
&=& m_{0} + m_{1}\,e^{2\pi i \ell/k} + 
m_{3}\,e^{-2\pi i \ell/k} 
- m_{2}\,\frac{\sin (3\pi\ell/k)}{\sin (\pi \ell/k)}\,,
\nonumber \\
\label{eigen1}
\end{eqnarray}
where in the last line $\ell \neq 0$. Putting in the values from 
Eq.~(\ref{ms}) gives
\begin{equation}
\lambda_{\ell}= \left\{ \begin{array}{ll} 
\beta + k\alpha, & \mbox{\ if $\ell = 0$} \\ 
\beta + 2ir\phi^{*} \sin( 2\pi\ell/k), & \mbox{\ if $\ell \neq 0$\,.}
\end{array} \right.
\label{eigen2}
\end{equation}
 
 \vspace{3 truecm}
 
%\newpage

\end{document}